\documentclass[12pt]{article}
\newlength{\dinwidth}
\newlength{\dinmargin}
\usepackage{amsmath}
\usepackage{amssymb}
\usepackage{latexsym}
\def\xxx#1 {{\sf hep-th/#1} }
\def\G{\Gamma}
\def\D{\Delta}

\def\a{\alpha}
\def\b{\beta}

\def\m{\mu}
\def\n{\nu}
\def\s{\sigma}

\def\t{\tau}
\def\o{\omega}

\def\mc{\mathcal}

\def\p{\partial}
\def\la{\langle}
\def\ra{\rangle}
\def\dg{\dagger}
\def\lra{\longrightarrow}

\newcommand{\ba}{\begin{array}}
\newcommand{\ea}{\end{array}}
\newcommand{\be}{\begin{equation}}
\newcommand{\ee}{\end{equation}}
\newcommand{\bea}{\begin{eqnarray}}
\newcommand{\eea}{\end{eqnarray}}

\def\thefootnote{\fnsymbol{footnote}}

\begin{document}
\thispagestyle{empty}
\addtocounter{page}{-1}
\begin{flushright}
\vskip-0.3cm
AEI-2002-016 \\
SNUST 02-0301\\
{\tt hep-th/0203080}
\end{flushright}
\vspace*{-0.5cm}
\centerline{\Large \bf Superstring on PP-Wave Orbifold}
\vspace*{0.3cm}
\centerline{\Large \bf from}
\vspace*{0.3cm}
\centerline{\Large \bf Large-N Quiver Gauge Theory
~\footnote{
Work supported in part by BK-21 Initiative in Physics (SNU - 
Project 2), KOSEF Interdisciplinary Research Grant 98-07-02-07-01-5, and 
KOSEF Leading Scientist Program.}}
\vspace*{1.3cm} 
\centerline{\bf Nakwoo Kim ${}^{a}$, Ari Pankiewicz ${}^a$, 
Soo-Jong Rey ${}^{b,c}$, Stefan Theisen ${}^a$}
\vspace*{0.8cm}
\begin{center}
\it Max-Planck-Institut f\"ur Gravitationsphysik, Albert-Einstein-Institut \\
Am M\"uhlenberg 1, D-14476 Golm, \rm GERMANY ${}^a$ \\
\vspace*{0.2cm}
\it School of Physics \& Center for Theoretical Physics \\
Seoul National University, Seoul 151-747, \rm KOREA ${}^b$\\
\vspace*{0.2cm}
\it Isaac Newton Institute for Mathematical Sciences \\
20 Clarkson Road, Cambridge CB3 0EH, \rm U.K. ${}^c$
\end{center}
\vspace*{0.7cm}
\centerline{\tt kim, apankie, theisen@aei-potsdam.mpg.de \hskip0.5cm  
sjrey@gravity.snu.ac.kr}
\vspace*{1.5cm}
\centerline{\bf abstract}
\vspace*{0.5cm}
\noindent
We extend the proposal of Berenstein, Maldacena and Nastase to Type IIB 
superstring propagating on pp-wave over ${\bf R}^4/{\bf Z}_k$ orbifold. 
We show that first-quantized free string theory is described correctly 
by the large $N$, fixed gauge coupling limit of ${\cal N}=2$ $[U(N)]^k$ 
quiver gauge theory. We propose a precise map between gauge theory operators 
and string states for both untwisted and twisted sectors. We also compute 
leading-order perturbative correction to the anomalous dimensions of these 
operators. The result is in agreement with the value deduced from string 
energy spectrum, thus substantiating our proposed operator-state map. 

\vspace*{1.1cm}

\baselineskip=18pt
\setcounter{footnote}{0}
\def\thefootnote{\arabic{footnote}}
\newpage
\section{Introduction}
Berenstein, Maldacena and Nastase (BMN) \cite{bmn} have recently put forward a 
remarkable proposal, extending the regime of gauge theory description from
supergravity to the full-fledged closed string theory. The idea is to 
utilize conserved global charges in the gauge theory and reorganize
the perturbative expansion. For example, for ${\cal N}=4$ super
Yang-Mills theory in four-dimensions, correlation functions involving
operators of scaling dimension $\Delta$ and global charge $J$ are
reorganizable in a calculable manner in the limit:
\bea
\lambda^2 = g^2_{\rm YM} N \rightarrow \infty, 
\qquad 
g^2_{\rm eff} = \frac{\lambda^2}{J^2} \rightarrow {\rm finite},
\qquad
\left(\Delta - J \right) \rightarrow {\rm finite}.
\label{BMNlimit}
\eea
In this limit, effective expansion parameter is set by $g^2_{\rm
eff}$, in sharp contrast to the `t Hooft large-N limit, where the
expansion parameter is set by the `t Hooft coupling $\lambda^2$. 

In the ${\cal N}=4$ super Yang-Mills theory, operators with $(\Delta - J) 
> 0$ are long supermultiplets, hence, correspond to massive string oscillator 
modes of Type IIB string theory. In the new limit Eq.(\ref{BMNlimit}), 
$J \sim \sqrt{N}$ is large of order 
${\cal O}\left(1 \slash \sqrt{g_{\rm s}} \right)$ 
at weak string coupling limit. The limit Eq.(\ref{BMNlimit}) turns out to 
correspond in the Type IIB string theory to the so-called Penrose limit of 
the $AdS_5 \times S^5$ background, yielding pp-wave spacetime with 
transverse ${\bf R}^4 \times {\bf R}^4$ geometry and homogeneous RR 5-form 
field strength \cite{papa1}--\cite{papa3}. 
It amounts to boosting the background around a great circle in $S^5$ 
and rescaling so that a neighborhood around null geodesics is blown up. 
As such, the BMN limit of ${\cal N}=4$ super Yang-Mills theory is interpretable
as gauge theory description for discrete light-cone quantization of Type IIB 
superstring. Interestingly, in the Penrose limit of $AdS_5 \times S^5$, the 
total number of isometries as well as spacetime supersymmetries remain the 
same. Rather, the limit yields a contraction of the $SU(2,2 \vert 4)$ 
superconformal algebra (see \cite{contraction} for an explicit demonstration).

A central technical feature that facilitates this correspondence is the
phenomenon that anomalous dimensions of a certain class of long multiplet 
operators are parametrically suppressed. Relevance of this sort of operators
to the string theory has been first emphasized by Polyakov \cite{polyakov}. 
In the proposal of BMN, these operators play the prominent role in that they
describe the creation and annihilation operators of string oscillation modes. 

An immediate question is whether the proposal is applicable to a more
nontrivial background of the pp-wave front. In this
paper, we extend the BMN proposal to the simplest yet nontrivial 
situation: pp-wave orbifold --- the homogeneous pp-wave background
(part of) whose transverse space is orbifolded. Specifically, we will
consider orbifolding one of the two ${\bf R}^4$ subspaces
transverse to the propagation null vector. 
Our motivation comes from various corners. First, the plane-wave
background considered by BMN preserves all 32 spacetime
supersymmetries. It is clearly of interest to investigate if the BMN
proposal is extendible to plane-wave backgrounds with fewer number of 
spacetime supersymmetries. The simplest way to reduce the supersymmetry is 
to orbifold part of the transverse space. Second, as shown in \cite{bmn, metsaev, mt}, 
the plane-wave background acts as a harmonic potential
to the string, and hence dynamical distinction between untwisted and twisted
states is less clear. It is thus of intrinsic interest to see if one can 
find a precise map between Type IIB string oscillation modes and quiver gauge 
theory operators, both for untwisted and twisted sectors. 

This paper is organized as follows. In section 2, we study the discrete 
light-cone quantization of the Type IIB superstring on pp-wave orbifold, and 
obtain the energy spectrum. In
section 3, we analyze gauge-invariant operators in the dual, ${\cal N}=2$ 
quiver gauge theory, and find precise correspondence with the spectrum 
obtained in section 2. In section 4, we compute perturbatively the anomalous 
dimension of $(\Delta - J) = 1$ operators at leading order and find agreement
with the light-cone energy spectrum of section 2.     

Shortly after \cite{bmn}, several preprints with various generalizations
have appeared \cite{IKM,GO,cobi}. In particular, \cite{ASJ}, which has 
substantial overlap with part of our work, was posted on the archive while we 
were in the process of writing up our results. 

\section{Type IIB Superstring on pp-wave Orbifold}
Begin with dynamics of a Type IIB superstring on pp-wave background. 
The background, supported by homogeneous RR 5-form and dilaton fields,
is given by 
\begin{align}
ds^2 & =-4dx^+dx^--\m^2(\vec{x}^2+\vec{y}^2)(dx^+)^2+d\vec{x}^2+d\vec{y}^2\,,\label{pp1}\\
F_{+1234} & =F_{+5678} = \m, \\
e^\phi &\equiv g_s = {\rm constant},
\label{pp2}
\end{align}
where $(\vec{x},\vec{y})\in{\bf R}^4 \times {\bf R}^4$, and is known to be maximally
supersymmetric, preserving all 32 spacetime supersymmetries. 
It was argued that, in the background Eqns.\eqref{pp1}--\eqref{pp2}, the Type IIB superstring
is exactly solvable \cite{metsaev,mt}, owing mainly to the fact that the light-cone worldsheet
dynamics is described by {\sl free} fields, albeit being massive. 

Recently, it was found that the pp-wave background Eqns.(\ref{pp1})--(\ref{pp2}) is 
related to the other known maximally
supersymmetric background -- $AdS_5\times S^5$ with RR 5-form 
flux threaded on the five-sphere -- via the Penrose limit along a large
circle of the $S^5$ \cite{papa2}. Note that the isometry group of 
the eight-dimensional space transverse to the null propagation direction is 
$SO(4)\times SO(4)$: while
the spacetime geometry is invariant under $SO(8)$, the 5-form field
strength breaks it to $SO(4)\times SO(4)$. In the Green-Schwarz
action of the Type IIB string in the plane-wave background, the
reduction of the isometry is due to the coupling of spinor fields to
the background RR 5-form field strength.

One is interested in reducing the number of supersymmetries preserved
by the background. As alluded to in the introduction, one can reduce the
32 supersymmetries to 16 supersymmetries by taking a ${\bf Z}_k$ orbifold 
of the ${\bf R}^4$ subspace parametrized by $\vec{y}$. 
The orbifold action is defined by
\begin{equation}
g : \quad (z^1,z^2) \longrightarrow (\o z^1,\o z^2)\qquad
{\rm where} \qquad \o=e^{\frac{2\pi i}{k}}\,.
\end{equation}
Here, $z^1\equiv\frac{1}{\sqrt{2}}(y^6+iy^7)$, 
$z^2\equiv\frac{1}{\sqrt{2}}(y^8-iy^9)$. The orbifold action $g$ acts on 
spacetime fields as $g=\exp \left( \frac{2\pi i}{k}(J_{67}-J_{89})\right)$, 
$J_{67}$ and $J_{89}$ being the rotation generators in the $67$ and $89$ 
planes, respectively. Defined so, the orbifold of
the pp-wave background is actually derivable from the Penrose limit of 
$AdS_5\times S^5/{{\bf Z}_k}$ taken along the great circle of the $S^5$ that 
is fixed by the orbifold.

In the light-cone gauge, Type IIB superstring on the pp-wave background 
Eqns.(\ref{pp1})--(\ref{pp2}) is described by eight worldsheet scalars $x^I$ and eight worldsheet 
Majorana fermions $(\theta_1,\theta_2)$, all of which are free but massive. 
The masses of the scalars and the fermions are equal by worldsheet supersymmetry (which descends from the light-cone gauge fixing of the Green-Schwarz action)
and equal the RR 5-form field strength, $\m$. 
Both $\theta_1,\theta_2$ are positive chirality Majorana-Weyl spinors of 
$SO(9,1)$, obeying the light-cone gauge condition $\G^+\theta_i=0$. 
Decomposing the worldsheet fields into $SO(4)_1\times SO(4)_2$ subgroups, 
\begin{align}
x^I & =(\vec{x},\vec{y})  \to (\vec{x},z^1, z^2)\,,\qquad\quad\qquad 
g : \quad 
\vec{x} \longrightarrow \vec{x}\,,\quad 
z^m \longrightarrow \o z^m\,,\\
\theta &\equiv\frac{1}{\sqrt{2}}(\theta_1+i\theta_2)  \to (\chi^{\a},\xi^{\dot{\a}})\,,\qquad
g : \quad 
\chi^{\a} \longrightarrow \chi^{\a}\,, \quad 
\xi^{\dot{\a}} \longrightarrow {\Omega^{\dot{\a}}}_{\dot{\b}}\xi^{\dot{\b}}\,.
\end{align}
Here, $\a$ and $\dot{\a}$ are spinor indices of $SO(4)_2$, ranging over
1,2. We have suppressed the spinor indices of $SO(4)_1$ under 
which $\chi^{\a}$ carry positive chirality, while $\xi^{\dot{\a}}$ carry 
negative one. $\Omega={\rm diag} (\o,\o^{-1})$, viz. $\xi^{\dot{1}}$ and 
$\xi^{\dot{2}}$ transform oppositely under the ${\bf Z}_k$ orbifold action. 
It is convenient to combine $\xi^{\dot{1}}, \overline{\xi}^{\,\dot{2}}$ into a Dirac spinor $\xi$, 
and $\overline{\xi}^{\,\dot{1}}$ and $\xi^{\dot{2}}$ into its conjugate $\overline{\xi}$ and analogously
for $\chi$ and $\overline{\chi}$. 
As the worldsheet theory is free, it is straightforward to quantize the 
Type IIB superstring in each twisted sector, the only difference among
various sectors being the monodromy of the worldsheet fields sensitive to 
the orbifolding, viz. $z^1, z^2$ and  $\xi$. The other worldsheet 
fields remain periodic as usual. 
The monodromy conditions in the $q$-th twisted sector,
$q=0,\ldots,k-1$, are given by 
\begin{equation}
z^m(\s+2\pi\a' p^+,\t)=\o^qz^m(\s,\t)\,,\qquad
\xi (\s+2\pi\a' p^+,\t)=\o^q\xi (\s,\t)\, ,
\end{equation}
and result in fractional moding, $n(q)=n+\frac{q}{k}$  $\left(n\in{\bf Z}
\right)$ of the corresponding oscillator modes.

Physical states are obtainable by applying the bosonic and fermionic creation 
operators to the light-cone vacuum $|0,p^{+}\ra_q$ of each $q$-th 
twisted sector. They ought to satisfy additional constraints ensuring the 
level-matching condition: 
\begin{equation}
\sum_{n\in{\bf Z}}nN_n=0, \quad
\sum_{n\in{\bf Z}}n(q)\left(N_{n(q)}-\overline{N}_{-n(q)}\right)
=\sum_{n\in{\bf Z}}\left(n(q)N_{n(q)}+n(-q)\overline{N}_{n(-q)}\right) = 0,
\label{levelmatching}
\end{equation}
and ${\bf Z}_k$ invariance. The bosonic creation operators consist of
\begin{equation}
\vec{a}_n^{\dg}\,,\qquad\text{ and }\qquad\a^{\dg\,m}_{n(q)}\,,\quad \overline{\a}^{\dg\,m}_{n(-q)} \, \qquad \left( n \in {\bf Z} \right).
\end{equation}
Here, $\vec{a}_n$ are the $\vec{x}$ oscillators, whereas $\a^m_{n(q)}$ and 
$\overline{\a}^m_{n(-q)}$ are $z^m$ and $\overline{z}^m$ oscillators, respectively. 
The fermionic creation operators consist, in obvious notation, of
\begin{equation}
\chi^{\dg}_n\,,\quad \overline{\chi}^{\dg}_n\qquad\text{ and }\qquad
\xi^{\dg}_{n(q)}\,,\quad \overline{\xi}^{\dg}_{n(-q)}\,.
\end{equation}
Acting the fermionic zero mode oscillators to the light-cone vacua and 
projecting onto ${\bf Z}_k$ invariant states, one fills out 
${\mc N}=2$ gravity and tensor supermultiplets of the plane wave background. 
The action of the bosonic oscillators on these gives rise to a whole tower of 
multiplets, much as in the $AdS_5 \times S^5$ case. 
As an example, we have four invariant states 
with a single bosonic oscillator 
\begin{equation}
\vec{a}_0^{\,\dg}|0,p^+\ra_q\,,
\end{equation}
and states with two bosonic oscillators are
\begin{equation}\label{osc2}
a_n^{\dg\,\m}a_{-n}^{\dg\,\n}|0,p^{+}\ra_q\,,\qquad 
\a^{\dg\,l}_{n(q)}\overline{\a}^{\dg\,m}_{-n(q)}|0,p^{+}\ra_q\,.
\end{equation}
In the ${\bf Z}_2$ case there are additional invariant states 
built from two $z^m$ or two $\overline{z}^m$ oscillators. However, they do not 
satisfy the level matching condition, Eq.(\ref{levelmatching}).

One straightforwardly obtains the light-cone Hamiltonian in the 
$q$-th twisted sector as
\begin{equation}
H_{\text{LC}}(q)=
\sum_{n\in{\bf Z}}N_n\sqrt{\m^2+\frac{n^2}{(\a'p^+)^2}}+
\sum_{n\in{\bf Z}}\left(N_{n(q)}+\overline{N}_{-n(q)}\right)
\sqrt{\m^2+\frac{n(q)^2}{(\a'p^+)^2}}\,.
\label{lchamiltonian}
\end{equation}
The first sum is over those oscillators which are not sensitive to the 
orbifold. Positive modes label `left' movers, 
negative ones `right' movers, $N_n$ ($N_{n(q)}$ and $\overline{N}_{-n(q)}$) is the total occupation number of bosons and 
fermions. The ground state energy is cancelled 
between bosons and fermions. This corresponds to a choice of fermionic zero mode vacuum that explicitly breaks the 
$SO(8)$ symmetry, which is respected by the metric but not the field strength 
background, to $SO(4)_1\times SO(4)_2$ \cite{mt}. 
\section{Operator Analysis in ${\mc N}=2$ Quiver Gauge Theory}
It is known \cite{ks} that Type IIB supergravity on 
$AdS_5 \times (S^5 \slash {\bf Z}_k)$ is dual to 
${\mc N}=2$ $[U(N)]^k$ quiver gauge theory, the worldvolume theory of 
$kN$ D3-branes sitting at the orbifold singularity. 
In light of discussions in the previous 
section, one anticipates that Type IIB superstring on pp-wave orbifold is 
dual to a new perturbative expansion of the quiver gauge theory
at large $N$ and {\em fixed} gauge coupling $g^2_{\text{YM}} = 4 \pi g_s k$. 
The factor of $k$ in the relation between the string and the gauge coupling
is standard, and is easily deducible from moving the D3-branes off the tip of 
the orbifold into the Higgs branch. See also \cite{LNV}.
In the new expansion, one focuses primarily on states with conformal weight 
$\D$ and $U(1)_R$ charge $J$ which scale as $\D$, 
$J\sim\sqrt{N}$, whose difference $(\D-J)$ remains finite in the large 
$N$ limit. $U(1)_R$ is the subgroup of the orginal 
$SU(4)_R$ symmetry of ${\mc N}=4$ super Yang-Mills theory, which on the 
gravity side corresponds to the $S^1$ fixed under the orbifolding; 
this 
$U(1)_R$ together with the $SU(2)_1$ subgroup of the remaining $SO(4)\simeq SU(2)_1\times SU(2)_2$ that commutes 
with ${\bf Z}_k\subset SU(2)_2$ forms the $R$-symmetry group 
of ${\mc N}=2$ supersymmetric gauge theory. 

The reason for the above scaling behaviour is that 
$(\D-J)$ is identified with the light-cone Hamiltonian 
on the string theory side, whereas 
\footnote{
Since $\int_{S^5/{\bf Z}_k} F_5=N$, 
the radius of $AdS_5$ is proportional to $(kN)^{1/4}$.}
$\frac{J}{\sqrt{kN}}\sim p^+$, $p^+$ being 
the longitudinal momentum carried by the string. 
When $(\D-J) \ll J$,  the light-cone Hamiltonian Eq.(\ref{lchamiltonian})
implies that on the gauge theory side there are operators 
obeying the following relation between the dimension $\D$ and the 
$U(1)_R$ charge $J$ (we set $\m\equiv 1$)
\begin{equation}
(\D-J)_n=\sqrt{1+ g^2_{\rm eff} 
{n^2}}\,\qquad\text{ and }\qquad
(\D-J)_{n(q)}=\sqrt{1+
g^2_{\rm eff}
\left(n(q) \right)^2 }\,.
\label{lchamiltonian2}
\end{equation}

In the gauge theory, before orbifolding we have $N\times N$ matrix valued fields, i.e. the gauge field $A_{\m}$, 
complex scalars $Z=\frac{1}{\sqrt{2}}(X^4+iX^5)$ and 
$\phi^m=(\phi^1,\phi^2)\equiv\frac{1}{\sqrt{2}}(X^6+iX^7,X^8-iX^9)$, and 
fermions $\chi$ and $\xi$. The fields $\chi$ and $\xi$ are spinors of 
$SO(5,1)$, transforming as ${\bf 4}$ and ${\bf 4}'$, respectively.
For defining the ${\bf Z}_k$ orbifolding in the gauge theory, 
we promote these fields to $kN\times kN$ matrices 
${\mc A}_{\m}$, ${\mc Z}$, $\Phi^m$, ${\mc X}$ and $\Xi$ 
and project onto the ${\bf Z}_k$ invariant 
components. The projection is ensured by the conditions
\begin{equation}
S{\mc A}_{\m}S^{-1}={\mc A}_{\m}\,,\qquad S{\mc Z}S^{-1}={\mc Z}\,,\qquad 
S{\mc X}S^{-1}={\mc X}
\end{equation}
and 
\begin{equation}
S\Phi^mS^{-1}=\o\Phi^m\,,\qquad S\Xi S^{-1}=\o\Xi \,.
\end{equation}
where $S=\text{diag}(1,\o^{-1},\o^{-2},\ldots,\o^{-k+1})$, each block 
being proportional to the $N\times N$ unit matrix. 

The resulting spectrum is that of a four-dimensional 
${\cal N}=2$ quiver gauge theory \cite{dm} with $[U(N)]^k$ gauge group, 
containing hypermultiplets in the bi-fundamental representations of 
$U(N)_i\times U(N)_{i+1}$, $i\in{\bf Z}$ mod$(k)$. More precisely, 
${\mc A}_{\m}$, ${\mc Z}$ and ${\mc X}$ fill out $k$ ${\mc N}=2$ vector multiplets with the fermions transforming as 
doublets under $SU(2)_R$ 
(as its Cartan generator is proportional to $(J_{67}+J_{89})$). 
The ${\mc Z}$ field has the block-diagonal form
\begin{equation}
{\mc Z}=
\begin{pmatrix}
Z_1 &     &       &       &      & \\
    & Z_2 &       &       &      & \\
    &     &  Z_3  &       &      & \\
    &     &       &\quad\cdot&      & \\
    &     &       &       &\quad\cdot& \\
    &     &       &       &       & Z_k 
\end{pmatrix}
\end{equation}
with zeros on the off-diagonal and the diagonal blocks being 
$N\times N$ matrices of $U(N)_i$'s. 
The ${\mc A}_{\m}$ and ${\mc X}$ fields take an analogous form. Likewise,
the $\Phi^m$ and $\Xi$ fields fill out $k$ hypermultiplets, 
in which the scalars are doublets under $SU(2)_R$, 
whereas the fermions are neutral. 
The $\Phi^m$ fields take the form 
\begin{equation}
\Phi^m=
\begin{pmatrix}
0           & \phi^m_{12} &             &       &       &       & \\
            & 0           & \phi^m_{23} &       &       &       & \\
            &             & 0           & \quad\cdot &       &       & \\
            &             &             & \quad\cdot & \quad \cdot &     & \\
            &             &             &       &\quad\cdot & \quad\cdot & \\
\phi^m_{k1} &             &             &       &       &\quad\cdot&\quad\cdot
\end{pmatrix}
\end{equation}
and analogously for $\Xi$.

The light-cone vacua of the type IIB superstring in the 
plane-wave orbifold ought to be described by $p^-=0$. In the quiver gauge
theory side, the vacuum then corresponds to $(\D-J)=0$ 
operators acting on the Fock space vacuum.
What are the operators satisfying $(\D-J) =0$? 
Obviously, one can build $k$ mutually orthogonal, ${\bf Z}_k$ invariant 
single trace operators $\text{Tr}[S^q{\mc Z}^J]$. We propose that these 
operators are associated to the vacuum in the $q$-th twisted sector
\begin{equation}
\frac{1}{\sqrt{kJ}N^{J/2}}\text{Tr}[S^q{\mc Z}^J]\qquad\longleftrightarrow\qquad 
|0,p^+\ra_q\,,\qquad (q=0,\ldots,k-1)\,.
\end{equation}
In what sense is this identification unique? After all, in the quiver gauge
theory, it appears that the operators $\text{Tr}[S^q{\mc Z}^J]$ for 
any $q$ stand on equal footing. However, the orbifold action renders  
an additional ``quantum'' ${\bf Z}_k$ symmetry (see e.g. \cite{aps}) 
that acts on fields in the quiver gauge theory.\footnote{
This  ${\bf Z}_k$ should not to be confused with the spacetime ${\bf Z}_k$ 
used for constructing the orbifold. By construction, under the orbifold action,
all the fields are invariant.} 
Specifically, one can take an element $g$ in this quantum ${\bf Z}_k$, $g=e^{\frac{2\pi i}{k}}$, to act 
on an arbitrary field $T_{ij}$, $i,j\in{\bf Z}$ mod$(k)$, as 
$g: T_{ij} \lra T_{i+1,j+1}$. In particular, one notes that 
$g: \text{Tr}[S^q{\mc Z}^J] \lra \o^{q}\text{Tr}[S^q{\mc Z}^J]$. 
So one can indeed distinguish classes of operators on the quiver gauge
theory side by their eigenvalues under the quantum ${\bf Z}_k$ symmetry. 

Next, consider the eight twist invariant operators with $(\D-J)=1$. They are
\begin{align}
\frac{1}{kN^{(J+1)/2}}\text{Tr}[S^q{\mc Z}^J{\mc D}_{\m}{\mc Z}]
& \qquad\longleftrightarrow\qquad a_0^{\dg\,\m}|0,p^+\ra_q\,,\\
\frac{1}{kN^{(J+1)/2}}\text{Tr}[S^q{\mc Z}^J{\mc X}_{J=1/2}]
& \qquad\longleftrightarrow\qquad \chi^{\dg}_0|0,p^+\ra_q\,,\\
\frac{1}{kN^{(J+1)/2}}\text{Tr}[S^q{\mc Z}^J\overline{{\mc X}}_{J=1/2}]
& \qquad\longleftrightarrow\qquad \overline{\chi}^{\dg}_0|0,p^+\ra_q\,,
\end{align}
and hence identifiable with Type IIB supergravity modes 
(in each twisted sector) built out of a single zero-mode oscillator acting 
on the $q$-th vacuum. 
Here, ${\mc D}_{\m}{\mc Z}=\p_{\m}{\mc Z}+[{\mc A}_{\m},{\mc Z}]$ . 

Operators corresponding to higher string states on the pp-wave orbifold
arise as follows. Oscillators of non-zero level $n$
corresponding to the fields not sensitive to the orbifold 
are identified with insertions of the operators 
${\mc D}_{\m}{\mc Z}$, ${\mc X}_{J=1/2}$ and $\overline{{\mc X}}_{J=1/2}$
with a position dependent phase factor $e^{\frac{2\pi i l}{J}n}$ in the trace 
$\text{Tr}[S^q{\mc Z}^J]$. For instance, for $(\D - J) = 2$, 
\begin{equation}
\frac{1}{\sqrt{kJ}N^{J/2+1}}
\sum_{l=1}^J\text{Tr}[S^q{\mc Z}^l{\mc D}_{\m}{\mc Z}{\mc Z}^{J-l}{\mc D}_{\n}{\mc Z}]e^{\frac{2\pi i l}{J}n}
\qquad\longleftrightarrow\qquad a_n^{\m\,\dg}a_{-n}^{\n\,\dg}|0,p^{+}\ra_q\,.
\end{equation}
This is exactly the same as in the unorbifolded case -- 
the insertion of the position-dependent phase factor ensures 
that the level-matching condition is satisfied and that the light-cone 
energy of the string states is reproduced correctly \cite{bmn}. 

As for the remaining string states involving oscillators with a fractional 
moding $n(q)$ in the twisted sectors, we propose to identify them with 
insertions of the operators $\Phi^m$ and 
$\Xi_{J=1/2}$ together with the position-dependent phase factor 
$e^{\frac{2\pi i l}{J}n(q)}$. Similarly, insertions 
of $\overline{\Phi}^m$ and $\overline{\Xi}_{J=1/2}$ are accompanied with
the phase factor $e^{\frac{2\pi i l}{J}n(-q)}$. Again, the prescription 
implements the level-matching condition and, as will be demonstrated in the 
next section, seems 
to yield the correct energy of the corresponding string states. For instance, 
\begin{equation}\label{op}
\frac{1}{\sqrt{kJ}N^{J/2+1}}\sum_{l=1}^J\text{Tr}[S^q{\mc Z}^l\Phi^r{\mc Z}^{J-l}\overline{\Phi}^s]
e^{\frac{2\pi i l}{J}n(q)}
\qquad\longleftrightarrow\qquad \a^{r\,\dg}_{n(q)}\overline{\a}^{s\,\dg}_{-n(q)}|0,p^{+}\ra_q\,.
\end{equation}
Note that, for ${\bf Z}_2$ orbifold, the state 
$\sum_{k,l=1}^J\text{Tr}[S{\mc Z}^k\Phi^r{\mc Z}^{l}\Phi^s{\mc Z}^{J-k-l}]e^{\frac{2\pi i}{J}(kn(1)+(k+l)m(1))}$ corresponding to 
$\a^{r\,\dg}_{n(q)}\a^{s\,\dg}_{m(q)}|0,p^{+}\ra_1$, 
though being ${\bf Z}_2$ invariant, 
vanishes for all $m,n$ due to the cyclicity of the trace, as it should, cf. the remark below Eq.\eqref{osc2}.

Finally, operators with insertions such as ${\mc D}_{\m}{\mc D}_{\n}{\mc Z}$, $\overline{\mc Z}$ or ${\mc X}_{J=-1/2}$
in the trace are present at weak coupling, but should not be present at 
strong coupling, as there are no corresponding states in the string spectrum. 
As in \cite{bmn}, the reason for this might be related to the
fact that these operators acquire a large anomalous dimension in this 
limit \cite{bmn}. 
\section{Anomalously Suppressed Anomalous Dimensions}
In this section, in fixed $g^2_{\rm YM}$, large-$N$ and large-$J$ perturbation
theory, we shall be computing leading-order anomalous dimensions of 
$(\Delta - J) = 1$ operators in ${\cal N}= 2$ quiver gauge theory, and
confirm that our proposal for the twisted sector operators reproduces the
correct light-cone string energy spectrum. 
Amusingly, in the set-up we have outlined above, one can proceed
the computations essentially parallel to those of \cite{bmn}. 

The bosonic part of the (euclidean) quiver gauge theory action involving the 
transverse scalars is given by
\begin{align}\label{action}
S_{\text{YM}}=
\frac{2}{g^2_{\rm YM}}
\int d^4x\, &\text{Tr}  
\left[
\left| {\mc D} {\mc Z} \right|^2
+\sum_{m=1}^2 \left| {\mc D} \Phi^m \right|^2
-\frac{1}{2}\left(
\left| \left[ {\mc Z},\overline{\mc Z} \right] \right|^2
+\sum_{m=1}^2 
\left| \left[ \Phi^m,\overline{\Phi}^m \right] \right|^2
\right.
\right.
\nonumber\\
&
\left.\left.
+ \sum_{{m \neq n=1} 
}^2
\left( 
\left| \left[ \Phi^m,\overline{\Phi}^n \right] \right|^2
+\left| \left[ \Phi^m,\Phi^n \right] \right|^2 
\right)
+2\sum_{m=1}^2
\left(
\left| \left[ {\mc Z},\Phi^m \right] \right|^2
+\left| \left[ {\mc Z},\overline{\Phi}^m \right] \right|^2
\right)
\right) \right]\,.
\end{align}
The trace `Tr' runs over the $kN\times kN$ matrices, the 
$N\times N$ matrix blocks being invariant under the orbifold action. 

Explicitly, the quartic interactions involving ${\mc Z}$ with $\Phi^m$ 
(the last two terms in Eq.\eqref{action}) are 
\begin{align}
&-\sum_{a=1}^k\sum_{m=1}^2\frac{4}{g^2_{\rm YM}}
\int d^4x\,\text{tr}
\left[Z_a\phi^m_{a,a+1}\overline{Z}_{a+1}\overline{\phi}^m_{a+1,a}
+Z_a\overline{\phi}^m_{a,a-1}\overline{Z}_{a-1}\phi^m_{a-1,a}
\right]\nonumber\\
&+\sum_{a=1}^k\sum_{m=1}^2
\frac{2}{g^2_{\rm YM}}
\int d^4x\,\text{tr} 
\left[Z_a\phi^m_{a,a+1}\overline{\phi}^m_{a+1,a}\overline{Z}_a
+Z_a\overline{\phi}^m_{a,a-1}\phi^m_{a-1,a}\overline{Z}_a\right.\nonumber\\
&\qquad\qquad\qquad\left.
+\phi^m_{a,a+1}Z_{a+1}\overline{Z}_{a+1}\overline{\phi}^m_{a+1,a}
+\overline{\phi}^m_{a,a-1}Z_{a-1}\overline{Z}_{a-1}\phi^m_{a-1,a}
\right]\,,
\end{align}
the trace `tr' now being over $N\times N$ matrices of the $a$-th $U(N)$ group.
The first line contains `momentum-dependent' interactions, while the 
second and third line `momentum-independent' interactions, respectively.

The free field propagators are
\begin{equation}
\la\bigl(Z_a\bigr)_i^j(x)\bigl(\overline{Z}_b\bigr)_k^l(0)\ra=
\delta_{ab}\delta_i^l\delta_k^j
\frac{g^2_{\rm YM}}{8 \pi^2}
\frac{1}{|x|^2}\,,\quad
\la\bigl(\phi^m_{a,a+1}\bigr)_i^j(x)\bigl(\overline{\phi}^n_{b+1,b}\bigr)_k^l(0)\ra=
\delta_{ab}\delta^{mn}\delta_i^l\delta_k^j
\frac{g^2_{\rm YM}}{8 \pi^2}
\frac{1}{|x|^2}\,.
\end{equation}
If our proposed map between gauge theory operators and string modes
is correct, according to Eq.(\ref{lchamiltonian2}), anomalous dimension of 
the operators $O(x)$ in Eq.\eqref{op} is expected to 
receive perturbative corrections as 
\begin{equation}
(\D-J)_{n(q)}=1+
{1 \over 2} g^2_{\rm eff}
(n(q))^2 +\cdots.
\label{lchamiltonian3}
\end{equation}
We now demonstrate that this is precisely what one finds. 
At leading-order in perturbation theory, the logarithmically divergent 
contribution to the two-point function from the `momentum dependent' interactions
is obtained as
\begin{equation}
\label{twopt}
\la O(x)O^{\dg}(0)\ra=\frac{\left(
\frac{g^2_{\rm YM}}{8 \pi^2}
\right)^{\D}}{|x|^{2\D}}
\left[1+
\frac{g^2_{\rm YM} N}{2 \pi^2}
\cos\frac{2\pi n(q)}{J}
\ln\bigl(|x|\Lambda\bigl)\right]\,,
\end{equation}
whereas, for the `momentum-independent` interactions, 
$\cos\frac{2\pi n(q)}{J}$ is replaced by $-1$. 
Other `momentum-independent' interactions involving gauge bosons and scalar 
loops cancel, owing to the underlying ${\mc N}=2$ supersymmetry.
Hence, at large $J$ and $N$, the leading-order perturbative correction to the 
two-point correlation function is, up to overall normalization factor,
\begin{equation}
\la O(x)O^{\dg}(0)\ra\sim|x|^{-2\D}
\left[1- 
\frac{g^2_{\rm YM} N}{J^2}
(n(q))^2 \, \ln\bigl(|x|\Lambda\bigl)\right]\,.
\end{equation}
This implies that 
\begin{eqnarray}
(\D-J)_{n(q)}=1+
{1 \over 2} \frac{g^2_{\rm YM} N}{J^2}
\left(n(q) \right)^2+\cdots,
\end{eqnarray} 
reproducing precisely the anticipated perturbative correction  
Eq.(\ref{lchamiltonian3}), and hence the requisite light-cone energy
spectrum Eq.(\ref{lchamiltonian}) in the $q$-th twisted sector. 
Resummation of the leading logarithms, corresponding to multiple insertions 
of the above quartic interactions, is straightforward, and reproduces the 
full square-root form in Eq.(\ref{lchamiltonian2}). 
Again, this closely parallels the computation of \cite{bmn}.

\vskip1cm
\section*{Acknowledgement}
We thank Gleb Arutyunov for discussions. 
NK and SJR acknowledge hospitality of the Isaac Newton Institute for 
Mathematical Sciences during this work. AP and ST acknowledge support from GIF, the German-Israeli foundation for Scientific Research. NK acknowledges support from DFG.

\end{document}